# Fermi Surface and Pseudogap Evolution in a Cuprate Superconductor

Yang He[1], Yi Yin[1,a], M. Zech[1], Anjan Soumyanarayanan[1], Michael M. Yee[1], Tess Williams[1], M. C. Boyer[2,b], Kamalesh Chatterjee[2], W. D. Wise[2], I. Zeljkovic[1], Takeshi Kondo[3,c], T. Takeuchi[3], H. Ikuta[3], Peter Mistark[4], Robert S. Markiewicz[4], Arun Bansil[4], Subir Sachdev[1], E. W. Hudson[2,d,*], J. E. Hoffman[1,*]

[1] *Department of Physics, Harvard University, Cambridge, MA 02138, USA*
[2] *Department of Physics, Massachusetts Institute of Technology, Cambridge, MA 02139, USA*
[3] *Department of Crystalline Materials Science, Nagoya University, Nagoya 464-8603, Japan*
[4] *Department of Physics, Northeastern University, Boston, MA 02115, USA*
[a] *Current address: Department of Physics, Zhejiang University, Hangzhou, China*
[b] *Current address: Department of Physics, Clark University, Worcester, MA 01610, USA*
[c] *Current address: Institute for Solid State Physics, University of Tokyo, Japan*
[d] *Current address: Department of Physics, Pennsylvania State University, University Park, PA 16802, USA*
[*] *To whom correspondence should be addressed: jhoffman@physics.harvard.edu, ehudson@psu.edu*

The unclear relationship between cuprate superconductivity and the pseudogap state remains an impediment to understanding the high transition temperature ($T_c$) superconducting mechanism. Here we employ magnetic-field-dependent scanning tunneling microscopy to provide phase-sensitive proof that *d*-wave superconductivity coexists with the pseudogap on the antinodal Fermi surface of an overdoped cuprate. Furthermore, by tracking the hole doping (*p*) dependence of the quasiparticle interference pattern within a single Bi-based cuprate family, we observe a Fermi surface reconstruction slightly below optimal doping, indicating a zero-field quantum phase transition in notable proximity to the maximum superconducting $T_c$. Surprisingly, this **major reorganization of the system's underlying electronic structure** has no effect on the smoothly evolving pseudogap.

Superconductivity is one of several phenomena, including the pseudogap, that arises from interactions of electrons near the Fermi surface (FS) in hole-doped cuprates. The FS topology is therefore crucial to understanding these phenomena and their relationships. High-field quantum oscillation (QO) measurements (*1–3*) revealed a surprisingly small FS in underdoped $YBa_2Cu_3O_{6.5}$ (YBCO), in contrast to the conventional, large FS of overdoped cuprates like $Tl_2Ba_2CuO_{6+x}$ (*4*). Further high-field investigations led to the discovery of a quantum phase transition (QPT) at the low doping edge of this small FS regime, perhaps associated with a metal-insulator transition (*5*) or the formation of density-wave order (*6*). However, the large to small FS transition presumed to occur at higher doping has thus far not been observed by QO within a single hole-doped material system. Furthermore, it is unclear whether the small FS is merely revealed by QO or possibly created by the necessarily high magnetic fields.

A zero field alternative to QO, angle resolved photoemission spectroscopy (ARPES), has a long history of mapping the FS in $Bi_2Sr_2CaCu_2O_{8+x}$ (Bi2212) (*2, 7–11*). Here the onset of the pseudogap is defined by the opening of an anti-nodal gap and the reduction of the large FS to a "Fermi arc," which may actually be one side of a Fermi pocket, consistent with QO results (*10*). The pseudogap onset may be associated with a QPT just above optimal doping at $p$ = 0.19 (*11*). A second QPT to another competing phase is suggested to occur (*11*) at lower doping ($p$ = 0.076), similar perhaps to that found by QO (*5, 6*). However, if the transition near optimal doping is a FS reconstruction to pockets, as suggested in (*10*), why are sharp antinodal quasiparticles seen below this doping, all the way down to $p$ = 0.08 (*7, 9, 11*)? Further, if the antinodal FS persists down to $p$ = 0.08, what impact does the QPT associated with the onset of the pseudogap at $p$ = 0.19 have on the Fermi surface?

To address these outstanding questions we use scanning tunneling microscopy (STM) to study $(Bi,Pb)_2(Sr,La)_2CuO_{6+\delta}$ (Bi2201). In this hole-doped cuprate, the absence of bilayer splitting and the suppression of the supermodulation by Pb doping both simplify momentum space measurements. Additionally, the separation between the expected QPT near optimal doping and the well-characterized pseudogap onset at much higher doping (*12, 13*) allows clear investigation of the relationship between these two phenomena. We thus carry out a systematic investigation of Bi2201 at four different dopings (Fig. 1A), all of which display signatures of both the pseudogap and superconducting gap (*14, 15*) in their low temperature spectra (Fig. 1B). To extract FS information, we map their differential conductance $g(r, E)$ – proportional to the local density of states (DOS) of the sample – as a function of position $r$ and energy $E$. We use ratio maps, $Z(r, E) = g(r, +E) / g(r, -E)$, to enhance the QPI signal and cancel the setpoint effect (*16*), then we Fourier transform the data. This technique can highlight dominant wavevectors which arise from elastic scattering – quasiparticle interference (QPI) (*17, 18*) – between momentum space regions of high joint density of states (JDOS), thus enabling a probe of the FS.

In our most underdoped samples (UD25K and UD32K) we find a set of energy-dependent wavevectors $\{q_i\}$ following the "octet model" (*17*) (Fig. 2A). From these $\{q_i\}$, we extract points on the Fermi surface, but we find that they extend only to the antiferromagnetic Brillouin zone (AFBZ) boundary (Fig. 2I), similar to the behavior previously observed in Bi2212 (*18*). At higher doping (in UD32K, OPT35K, and OD15K samples) we find a "triplet" feature (outlined in black in Figs. 2F-H). This feature exists at multiple sub-pseudogap energies without apparent dispersion (Fig. S3). However, its relative prominence at large $q$ and low energies distinguishes it from the static checkerboard (*19, 20*), CDW (*21*), smectic (*22*), or fluctuating stripe (*23*) states which appear near $q \sim (0, \frac{1}{4} \cdot \frac{2\pi}{a_0})$ at low energies or near $q \sim (0, \frac{3}{4} \cdot \frac{2\pi}{a_0})$ around the pseudogap energy.

To understand the momentum space origin of this "triplet" QPI we follow Refs. (*24, 25*), and, taking into account reported spectral broadening (*9, 26*), particularly acute in the anti-node, compute the autocorrelation of all antinodal states within a small energy window of the high temperature FS. This autocorrelation corresponds to the low energy antinodal JDOS in the presence of pair breaking, and matches well with the data (compare Figs. 2K-L to Figs. 2G-H). The "triplet" feature thus arises from states near the antinodal FS. It could be considered as a continuation of "octet" QPI and we refer to it hereafter as "antinodal QPI." We conclude that the extinction of octet QPI at the AFBZ boundary in underdoped Bi2201 (Fig. 2I), followed by the appearance of antinodal QPI at higher doping, reveals the FS reconstruction (*1–4*) shown schematically in Fig. 2J.

Our UD32K sample shows both octet and antinodal QPI (Fig. 2F), suggesting that the QPT occurs near $p \sim 0.15$ doping. We support this point by using a modified Luttinger count to compute the hole concentration $p$ in both the large and small FS scenarios according to

$$p_{\text{large}} = \frac{2A_{\text{blue}}}{A_{\text{BZ}}} - 1; \quad p_{\text{small}} = \frac{A_{\text{pink}}}{A_{\text{AFBZ}}} = \frac{2A_{\text{pink}}}{A_{\text{BZ}}}$$

where $A_{\text{blue}}$ and $A_{\text{pink}}$ are the areas schematically indicated in the insets to Fig. 2J, and computed according to the tight binding fits in Table S2. The *x*-axis hole concentration is independently estimated from the measured $T_{\text{c}}$ (*27*). In either small or large FS scenario alone, we observe a sudden drop in Luttinger hole count from UD32K to UD25K. However, using $p_{\text{small}}$ in the UD25K sample and $p_{\text{large}}$ in the other samples, we find the expected linear relationship between the Luttinger count and the estimated hole concentration, providing further evidence of a small to large FS reconstruction upon increasing doping. Interestingly, these results agree well with high field Hall measurements of similar Bi2201 samples(*28*), suggesting that high field measurements in general, are revealing rather than creating small Fermi pockets.

Surprisingly, the pseudogap appears to be unaffected by this QPT, with a spectral signature that evolves smoothly through the transition. The pseudogap in Bi2201 is known (*12–14*) to exist well into the overdoped region of the phase diagram (Fig. 1A) rather than terminating near optimal doping, as in some other cuprates (*29*). Thus, the FS reconstruction we observe is distinct from the onset of the pseudogap.

We now face the question of how complete the coexistence of superconductivity and the pseudogap is, both in momentum- and real-space. To address the former we employ the phase-sensitive technique of magnetic-field-dependent QPI imaging (Fig. 3). In a superconductor, the QPI participants are Bogoliubov quasiparticles (BQPs), quantum-coherent mixtures of particles and holes. The scattering intensity of BQPs depends on the relative sign of the superconducting order parameter across the scattering wavevector. Furthermore, it is known empirically (*30*) and theoretically (*31*) that sign-preserving scattering will be enhanced by a magnetic field, while sign-reversing scattering will appear relatively suppressed. A *d*-wave superconducting order parameter changes sign in *k*-space (Fig. 3A); in our OD15K sample, the two branches of the Fermi surface around the antinode are close to each other (Fig. 2D), and are shown for simplicity in Fig. 3a as a single merged region at each antinode. Thus for a $d$-wave superconductor with antinodal BQPs, the simplified scattering process $q_A$ would be sign-preserving while $q_B$ would be sign-reversing. We observe that the sign-preserving $q_A$ scattering is enhanced in a magnetic field, while the sign-reversing $q_B$ scattering appears relatively suppressed (Fig. 3B; full field dependence in Fig. S12). This field-dependent QPI is thus fully consistent with the existence of $d$-wave BQPs in the antinodal FS.

For completeness, we consider several other possible causes of our field dependent antinodal QPI. First, we rule out a field-enhanced charge order, such as has been observed in vortex cores in Bi2212 (*19*), as it would not explain the existence of both enhanced and suppressed scattering. Second, we rule out $(\pi, \pi)$ and incommensurate orders such as antiferromagnetism, spin density waves, stripes, or $d$-density waves, which would not explain the features around both $(\pi, \pi)$ and $(2\pi, 0)$. Thus, we note that although non-phase-sensitive ARPES and STM (*12*, *32*) postulated a similar scenario, our field-dependent phase-sensitive QPI directly demonstrates the coexistence of $d$-wave BQPs and the pseudogap at the antinode.

Given that the pseudogap (PG) and superconductivity (SC) coexist in momentum space at the antinode, we next examine the nature of their spatial coexistence. We turn to a real space study of OD15K, where, amongst our samples, the pseudogap magnitude ($\Delta_{PG}$) is most comparable to the superconducting gap ($\Delta_{SC}$). To focus on the superconducting component, we suppress superconductivity with a 9 T field, and then remove the field-independent (PG) density of states by computing $S(r, E) = g(r, E, 0T) - g(r, E, 9T)$. Compared to Fig. 4A, where spatially disparate spectra show that $\Delta_{PG}$ varies by more than a factor of five across the field of view, $S(r, E)$ shows a relatively homogeneous gap of 6 mV (Fig. 4B), consistent with the homogeneous superconducting gap reported by a previous temperature dependent STM measurement on samples from the same batch (*14*). Our results thus support the two gap scenario (*8*, *14*, *32*, *33*) and suggest that $S(r, E)$ is indicative of the local superconducting spectrum.

Although the inhomogeneous pseudogap does not appear to relate to the magnitude $\Delta_{SC}$ of the superconducting order parameter (vertical dashed lines in Fig 4B), the coherence peak amplitude and gap depth, which have been shown to scale with superfluid density (*7*, *33*, *34*), decay markedly in regions of large $\Delta_{PG}$ (Fig. 4E). Similar to the ARPES analysis (*33*), we quantify the real space decoherence effect of the pseudogap by defining a local coherent spectral weight, $C_{\pm}(r) = S(r, \pm 6 \text{ mV}) - S(r, 0 \text{ mV})$, which increases with the height of the coherence peaks and depth of the gap. Visual comparison between maps of $\Delta_{PG}(r)$ and $C_+(r)$ (Figs. 4C and D), and their cross-correlation in Fig. 4F, demonstrate that stronger pseudogap (larger $\Delta_{PG}(r)$) correlates with local suppression of superconducting coherence on a very short (~2 nm) length scale.

Thus, our phase-sensitive momentum-space and normalized real-space measurements demonstrate that at high doping the pseudogap coexists with superconductivity in the antinode but correlates with suppressed superconducting coherence, suggesting a competitive relationship. We also find that the pseudogap transition is well separated from a zero field FS reconstruction that occurs near optimal doping – where superconductivity is strongest. The existence of a QPT near optimal doping has been long expected and suggested by a variety of other measurements (*29*). However, its differentiation from the pseudogap onset is surprising, and requires explanations of two phenomena rather than one. A number of theories have been proposed to explain the FS QPT. We can rule out the effect of crystal structure, based on its observed doping-independence (*35*). Our observed FS evolution is partly consistent with the phenomenological Yang-Rice-Zhang model (*36*), but does not show the completion of the arc to a pocket found in the closely related microscopic FL* model (*37*). Other orders such as antiferromagnetic fluctuations, charge order (*38*), *d*-density wave (*39*), quadrupole density wave (*40, 41*) and vestigial nematicity (*42*) are all consistent with our observation of the FS QPT. With respect to the PG onset, we observe that charge modulations are similarly unaffected by the FS QPT (Figs. S17 & S18), suggesting that the PG is associated with fluctuating charge order (*23*), which is pinned in all of our samples. Such charge order has recently been reported to compete with superconductivity in YBCO (*43*), while non-superconducting $La_{1.6-x}Nd_{0.4}Sr_xCuO_4$ shows static charge order ("stripes") (*44*), suggesting that such order (or its fluctuations) may be a universal competitor to superconductivity in the cuprates.

Figures

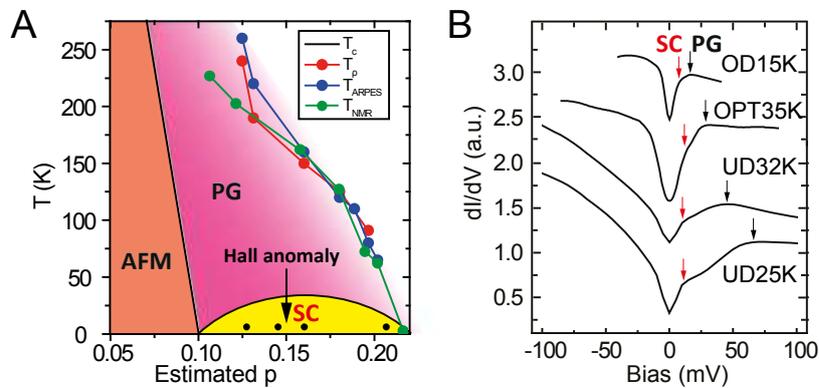

Fig. 1. *Phase diagram and spectra.* (A) Schematic temperature-doping phase diagram of Bi2201, showing antiferromagnetic insulator (AFM), superconductor (SC) and pseudogap (PG) phases. Four black points represent the sample batches of this study, namely underdoped UD25K and UD32K, optimal OPT35K, and overdoped OD15K. The pseudogap transition line *T\** is plotted as measured by ARPES (*12*), resistivity (*12*) and nuclear magnetic resonance (*13*). Anomaly in the Hall coefficient (*28*) is marked by the black arrow. (B) The spatially averaged differential conductance $g(E)$ for each sample. The pseudogap edge is marked with black arrows, whereas the low energy kink in each spectrum, considered to be related to the superconducting gap (*14, 15*), is marked with red arrows.

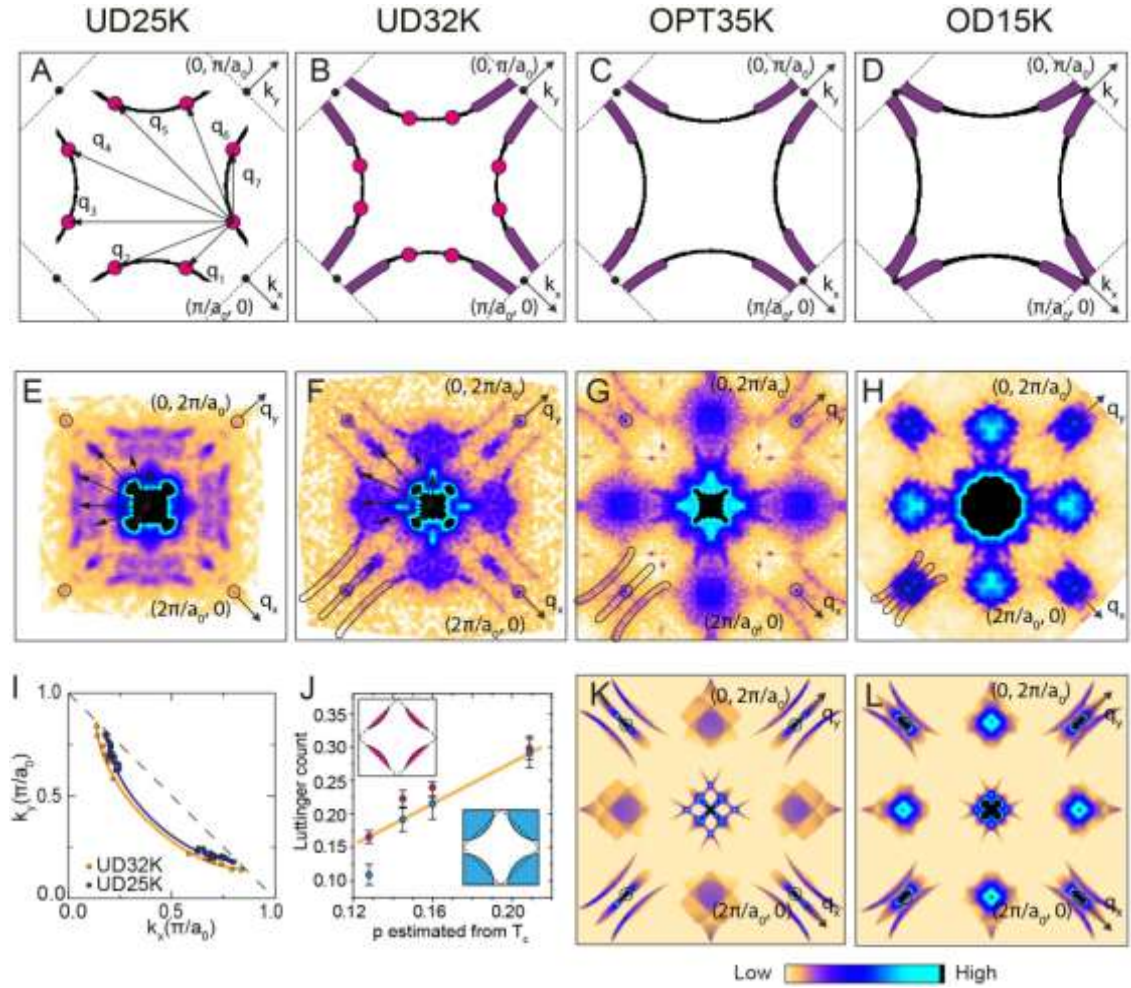

Fig. 2. Quasiparticle interference. (A-D) Schematic of the high DOS regions which contribute to QPI, superimposed on the calculated tight-binding FS (*15*) for each doping. Pink circles represent regions contributing to octet QPI. Purple regions represent the Fermi surface sections contributing to antinodal QPI. (E-H) Low energy $Z(q, E)$ on four samples: UD25K at $E$ = 9 mV (E), UD32K at $E$ = 5 mV (F), OPT35K at $E$ = 5 mV (G), and OD15K at $E$ = 3 mV (H). In UD25K (E) and UD32K (F) samples, conventional octet QPI is observed, dispersing with energy. In UD32K (F), OPT35K (G), and OD15K (H) samples, antinodal QPI appears. The evolution from octet to antinodal QPI shows the transition from small to large Fermi surface. (I) Locus of maximum DOS in the Bogoliubov band, from fitting octet QPI peaks. Lines show tight-binding fits. The BQP interference pattern vanishes around the AFBZ boundary (dashed line). (J) Luttinger hole counts from small FS (pink dots & upper inset) and large FS (blue dots & lower inset) scenarios vs. hole doping estimated from $T_c$. Error bar on the *y* axis comes from the 0.01 eV chemical potential uncertainty in the tight-binding parameterization. (K-L) Autocorrelation of broadened antinodal Fermi surface sections marked with purple in (C) and (D). This model of antinodal scattering agrees well with the data (G-H).

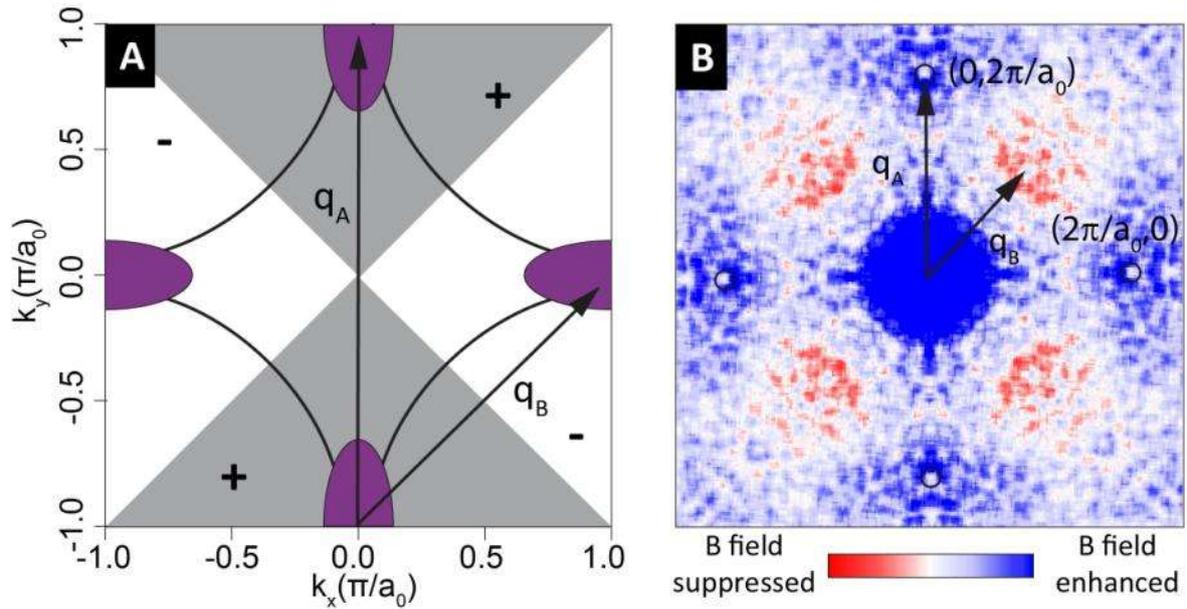

Fig. 3. Antinodal *d*-wave pairing. (A) First Brillouin zone of the OD15K sample, with calculated tight-binding FS. White and gray areas represent opposite signs of the *d*-wave superconducting order parameter. The antinodal bands approach each other in overdoped materials, and are represented by merged purple regions around *M* points for simplicity. The simplified scattering vectors $q_A$ and $q_B$ are shown as black arrows. (B) $Z(q, 6\text{ mV}, 9\text{T}) - Z(q, 6\text{ mV}, 0\text{T})$ showing *B*-suppressed (red) and *B*-enhanced (blue) weights, demonstrating the presence of superconducting Bogoliubov quasiparticles in the FS antinode where they coexist with the pseudogap.

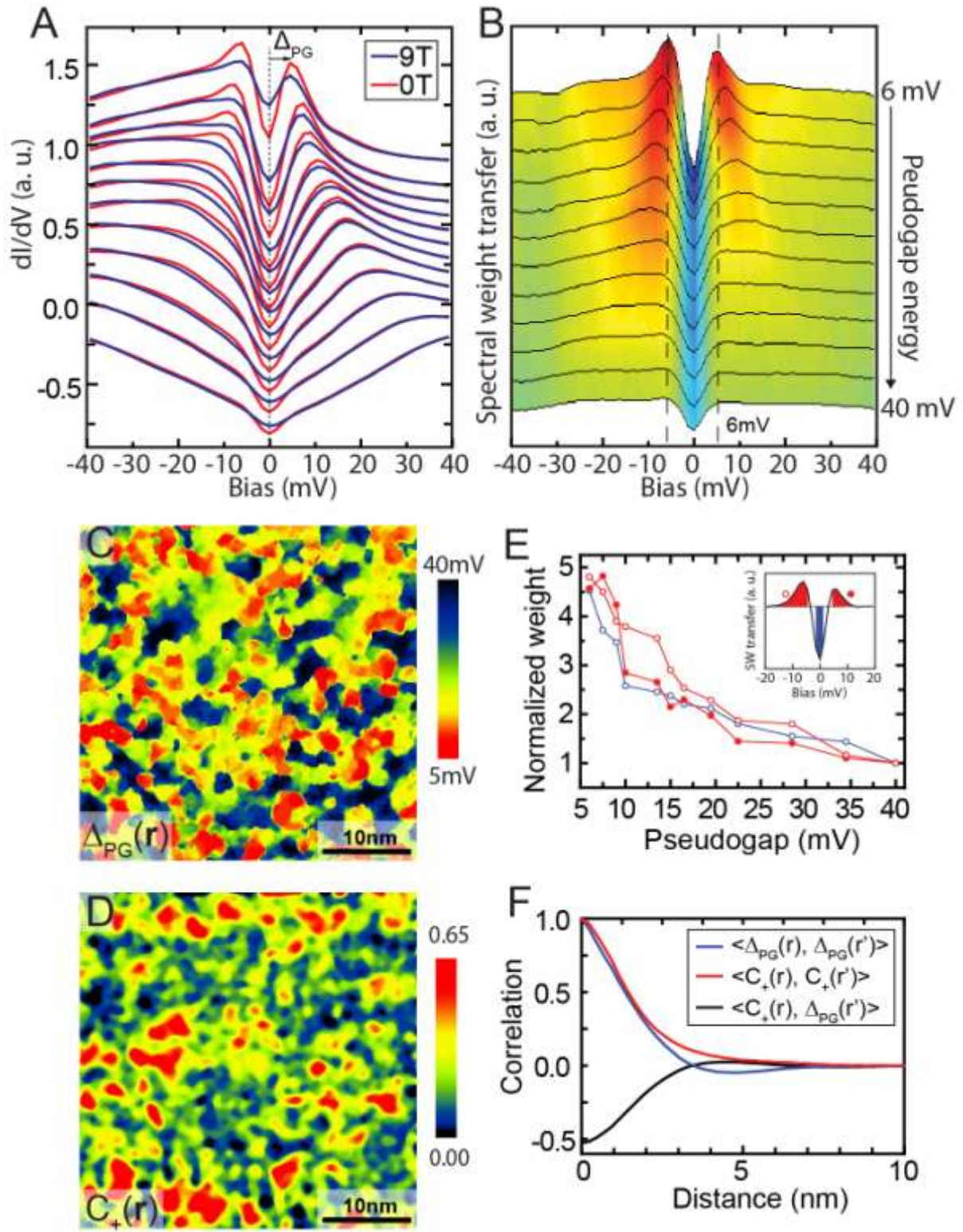

Fig. 4. Local pseudogap-induced decoherence. (A) OD15K spectra binned by pseudogap energy (at 0T), and spatially averaged within each bin at 0T (red) and 9T (blue). Pairs of spectra are offset for clarity. (B) Magnetic field induced spectral weight transfer, $S(r, E) = g(r, E, 0T) - g(r, E, 9T)$, for these same pairs. Because the PG is $B$-independent, $S(r, E)$ is attributed to superconductivity. Although the ~6 mV superconducting gap shows no trend as the PG increases more than five-fold, the superconducting coherence peak decreases monotonically with increasing $\Delta_{PG}$. (C) $\Delta_{PG}(r)$ in the same FOV from which Figs. 2h and 3b were calculated. (D) $C_+(r)$ map, defined as the difference between coherence peak amplitude and zero bias conductance: $S(r, 6\ mV) - S(r, 0\ mV)$, of the same area as (C). The color bar is reversed for better comparison to $\Delta_{PG}(r)$. (E) Dependence of coherence peak amplitude and gap depth on $\Delta_{PG}$. Coherence peak amplitude is integrated from ±6 mV to ±12 mV while gap depth is integrated from -1.5 mV to 1.5 mV. Red solid and hollow dots stand for the positive and negative coherence peak weight respectively. Blue hollow dots stand for gap depth. All data shown are normalized to their values at $\Delta_{PG}$ = 40 mV for better comparison. (F) Angle-averaged autocorrelations of $\Delta_{PG}(r)$ (blue) and $C_+(r)$ (red) and cross-correlation of $\Delta_{PG}(r)$ with $C_+(r)$ (black).

Acknowledgements
We thank Heon-Ick Ha for her role in the Harvard STM system construction. We thank Emanuele Dalla Torre, Eugene Demler, Steven Kivelson, Louis Taillefer, Philip Phillips, Fuchun Zhang, Ruihua He, J. C. Séamus Davis, Piers Coleman, Igor Mazin, Lara Benfatto, Chandra Varma, Tatsuya Honma, Mohit Randeria, and Nandini Trivedi for helpful discussion. Work at Harvard University was supported by the Air Force Office of Scientific Research under grant FA9550-06-1-0531, the U.S. National Science Foundation under grants DMR-0847433 and DMR-1103860, and the New York Community Trust – George Merck Fund. Work at MIT was supported in part by a Cottrell Scholarship awarded by the Research Corporation and by the MRSEC and CAREER programs of the NSF. Work at Northeastern University was supported by the US Department of Energy, Office of Science, Basic Energy Sciences contract number DE-FG02-07ER46352, and benefited from Northeastern University's Advanced Scientific Computation Center (ASCC), theory support at the Advanced Light Source, Berkeley and the allocation of time at the NERSC supercomputing center through DOE grant number DE-AC02-05CH11231. T.K., T.T., and H.I. synthesized and characterized the samples. M.B., K.C., and D.W. performed the STM measurements on UD25K, UD32K, and OPT35K samples; Y.Y., T.W. and M.Z. performed the magnetic-field-dependent STM measurements on OD15K samples. Y.H. analyzed the data and wrote the manuscript, with input from A.S., M.M.Y., and S.S. P.M. and R.S.M. performed theoretical simulations, with advice from A.B. E.W.H. and J.E.H. advised the experiments and analysis, and shaped the manuscript. The authors declare no competing financial interests. Correspondence and requests for materials should be addressed to J.E.H. (jhoffman@physics.harvard.edu) and E.W.H. (ehudson@psu.edu).

# Supplementary Information

# I. Experimental methods

We performed a systematic STM study on $(Pb_xBi_{2-x})La_ySr_{2-y}CuO_{6+z}$ (Bi2201), covering a large range of doping from UD25K to OD15K. Samples were grown with a floating zone technique. Pb doping was used to suppress the supermodulation. The data were acquired using two different home-built cryogenic STMs at a temperature of 6K. All samples were cleaved *in situ* in a cryogenic environment and inserted immediately into the STM. All tips were mechanically sharpened PtIr wires, cleaned by field emission and characterized on gold. Each dataset used in this study was acquired with atomic resolution, in a field of view (FOV) at least 40 x 40 nm$^2$, ensuring momentum precision better than 1% of $2\pi/a_0$. We used a lock-in amplifier to map the differential tunneling conductance $g(r,E)$, which is proportional to the local density of states. Data acquisition parameters used for the four samples are shown in Table S1.

Table S1 | Data acquisition parameters.

|  | UD25K | UD32K | OPT35K | OD15K |
|---|---|---|---|---|
| Setup V (mV) | -100 | -200 | -100 | -100 |
| Setup I (pA) | 400 | 400 | 400 | 100 |
| Bias Modulation (mV rms) | 5 | 5 | 5 | 2 |

An algorithm (*22*) was applied to remove the small thermal and piezoelectric drift which is inevitable during the acquisition of long maps. We use ratio maps, $Z(r,E) \equiv g(r,E)/g(r,-E)$, to enhance the QPI signal and cancel the STM set point effect (*18*). We Fourier transformed each energy layer of $Z(r,E)$ to obtain $Z(q,E)$, symmetrized (four-fold rotation and mirror) to increase signal-to-noise, and smoothed to remove visually distracting salt-and-pepper noise. Figure S1 exemplifies the detailed data processing procedure. Figure S2 compares the raw $g(q,E)$ to the ratio $Z(q,E)$; the qualitative similarity therein verifies that processing did not create artifacts.

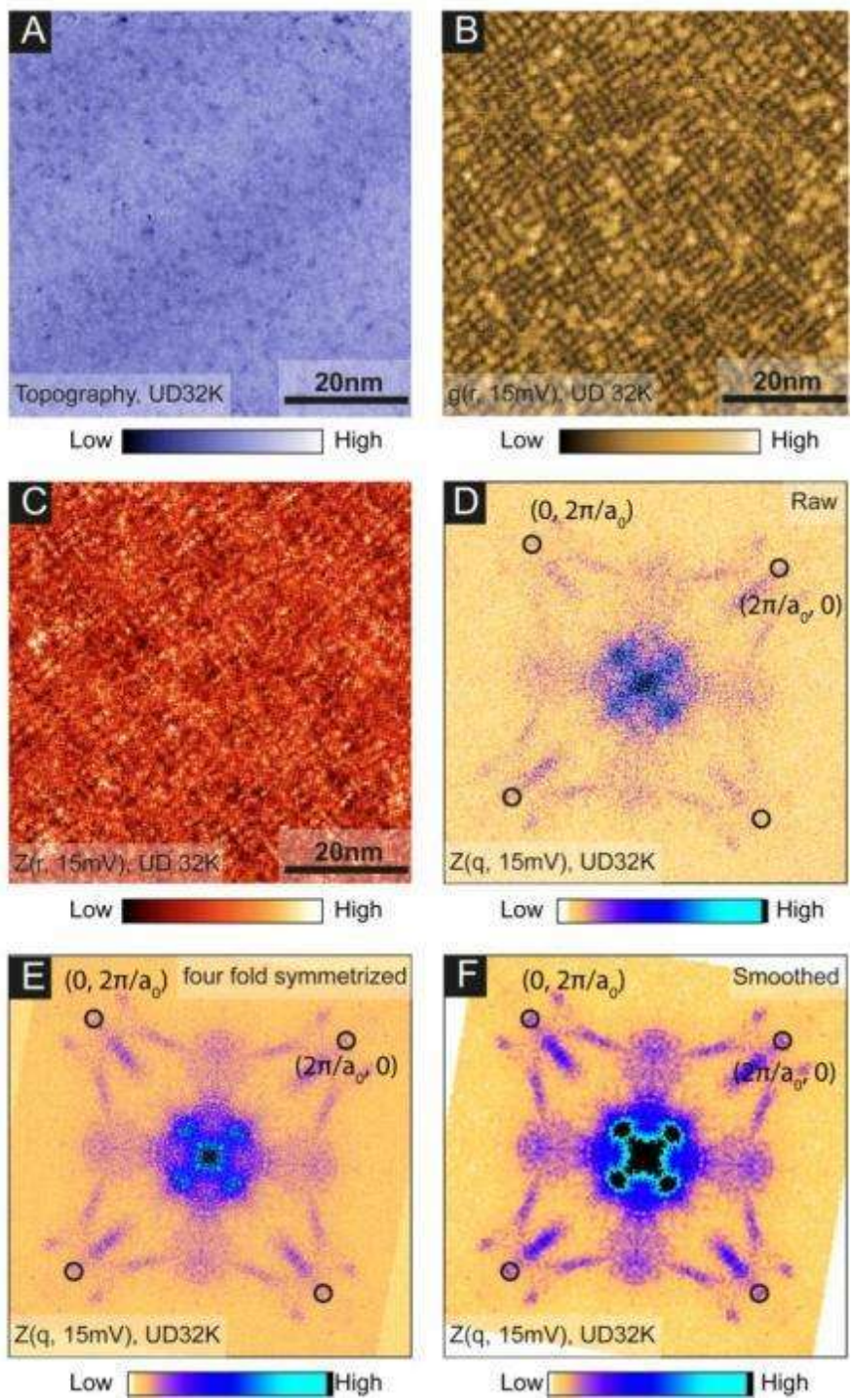

**Figure S1 | QPI analysis steps**
A, 66 nm × 66 nm topography of UD32K sample with atomic resolution. B, $g(r, 15\text{ mV})$ of the same FOV as in A. C, $Z(r, 15\text{ mV}) = g(r, 15\text{ mV})/g(r, -15\text{ mV})$. D, $Z(q, 15\text{ mV})$ is the Fourier transform of $Z(r, 15\text{ mV})$. Bragg peaks are highlighted with black circles. E, $Z(q, 15\text{ mV})$ is symmetrized (four-fold rotation & mirror), taking advantage of crystal symmetry to enhance signal to noise. F, $Z(q, 15\text{ mV})$ is smoothed to remove the salt-and-pepper noise and to enhance the main features.

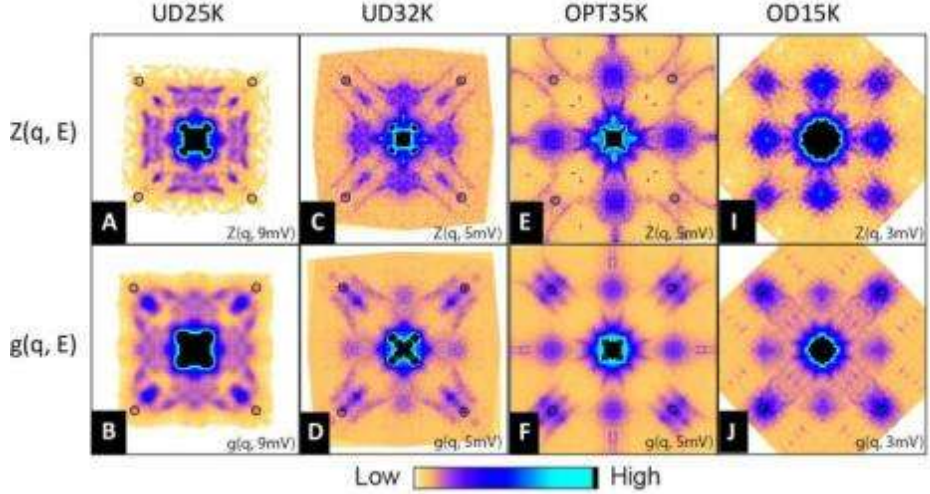

**Figure S2 | QPI raw data vs. ratio maps**
Comparison of raw $g(\mathbf{q}, E)$ data to ratio $Z(\mathbf{q}, E)$. In the UD25K and UD32K samples, octet QPI is enhanced in $Z(\mathbf{q}, E)$ as expected (*18*). In the UD32K, OPT35K, and OD15K samples, antinodal QPI is observed in both $g(\mathbf{q}, E)$ and $Z(\mathbf{q}, E)$.

## II. Quasiparticle Interference

QPI energy dependence is shown for all four samples in Figure S3. In the most underdoped UD25K sample, we observe only conventional octet QPI with clear dispersion, while in the optimal and overdoped OPT35K and OD15K samples, we observe only antinodal QPI. In the lightly underdoped UD32K sample we observe both. The absence of octet QPI in the OPT35K and OD15K samples is not yet understood. One possibility is that the octet QPI at these doping levels merges into the antinodal QPI, so that we are unable to distinguish between the two patterns. A second possibility is that increased broadening in these samples makes the effective Fermi arc long enough to connect to the antinodal region where pseudogap-induced decoherence prevents the observation of octet dispersion (*45*). A third possibility is that low-energy (near-unitary) scatterers necessary to observe octet QPI are present only in the underdoped samples.

We illustrate the octet QPI and its extinction at the antiferromagnetic Brillouin zone (AFBZ) boundary in Figure S4. In a superconductor, Bogoliubov quasiparticles (BQPs) disperse as $E(k) = \sqrt{\varepsilon(k)^2 + \Delta(k)^2}$ where $\varepsilon(k)$ is the normal state dispersion relation and $\Delta(k)$ is the superconducting order parameter. The gap $\Delta(k) = \Delta_0 \cos(2\theta_k)$ follows $d_{x^2-y^2}$ symmetry, i.e. it vanishes in the $(\pm\pi, \pm\pi)$ (nodal) directions and is maximized in the $(\pm\pi, 0)$ and $(0, \pm\pi)$ (antinodal) directions. The quasiparticle dispersion $E(k)$ thus gives rise to four "banana"-shaped contours of constant energy, depicted in Figure S4A. At each energy, the BQP density of states (DOS) is maximized at the eight "banana" tips, due to the large value of $1/\nabla_k(E)$ in those regions. Elastic scattering between these eight high-DOS regions gives rise to the dominant QPI vectors shown. Figure S4B emphasizes $q_1$ and $q_5$ which lie along Bragg directions, and their relation to the AFBZ. Within the octet model, $q_5$ would be expected to disperse all the way to $2\pi/a_0$ at energy $\Delta_0$. However, we show in Figure S4C-F that in both the UD25K and UD32K samples, $q_5$ disperses according to the octet model at low energies, then saturates at a static value $q_5^* < 2\pi/a_0$ at larger energies. The doping dependence of $q_1^*$ and $q_5^*$ have been previously reported and discussed (*21, 46, 47*).

In Figure S5 we contrast antinodal QPI, apparent in the UD32K, OPT35K, and OD15K samples, with its absence in the UD25K sample.

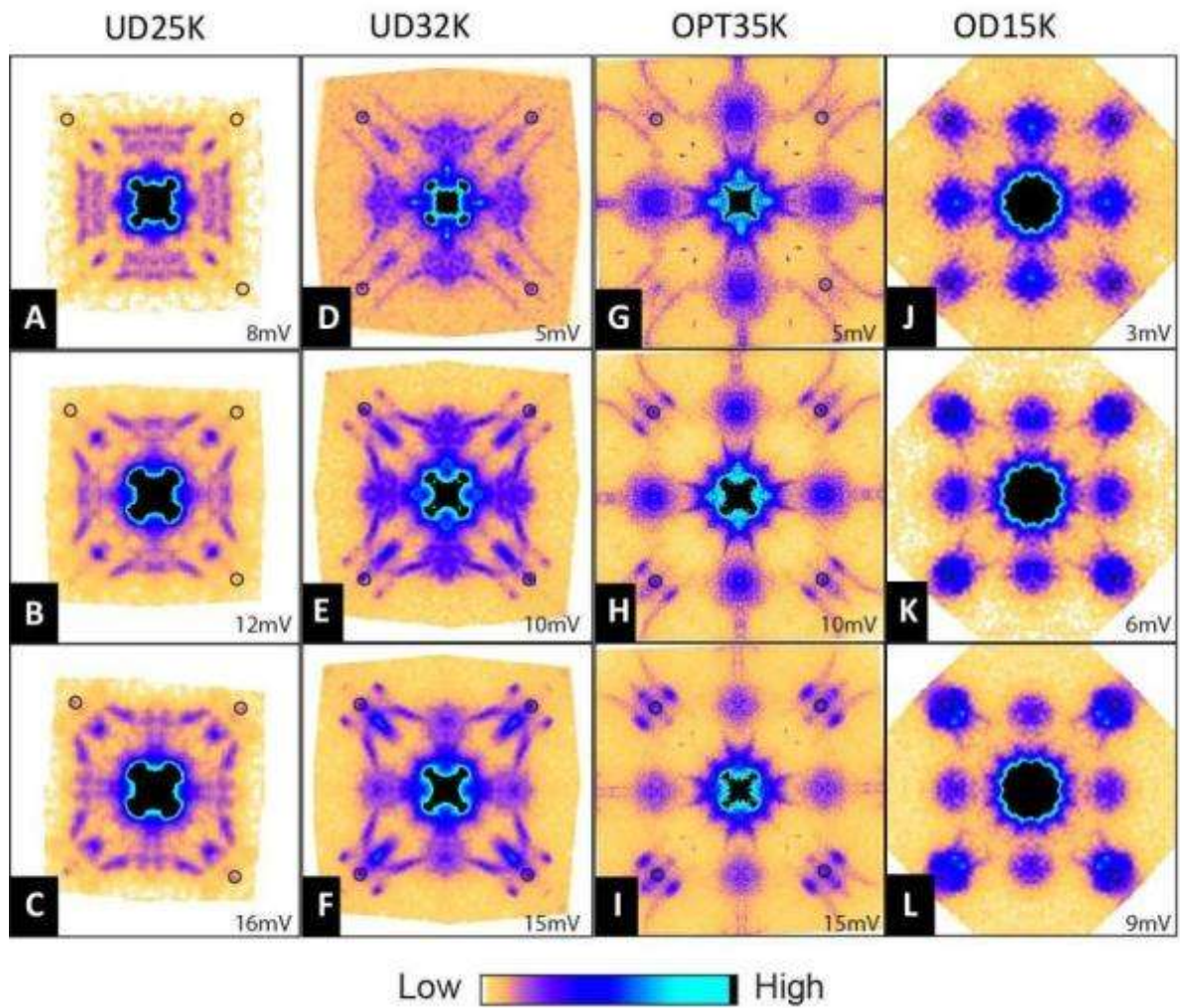

**Figure S3 | QPI energy dependence**
$Z(\mathbf{q}, E)$ is shown for three representative energies in each sample. Black circles show the Bragg peaks.

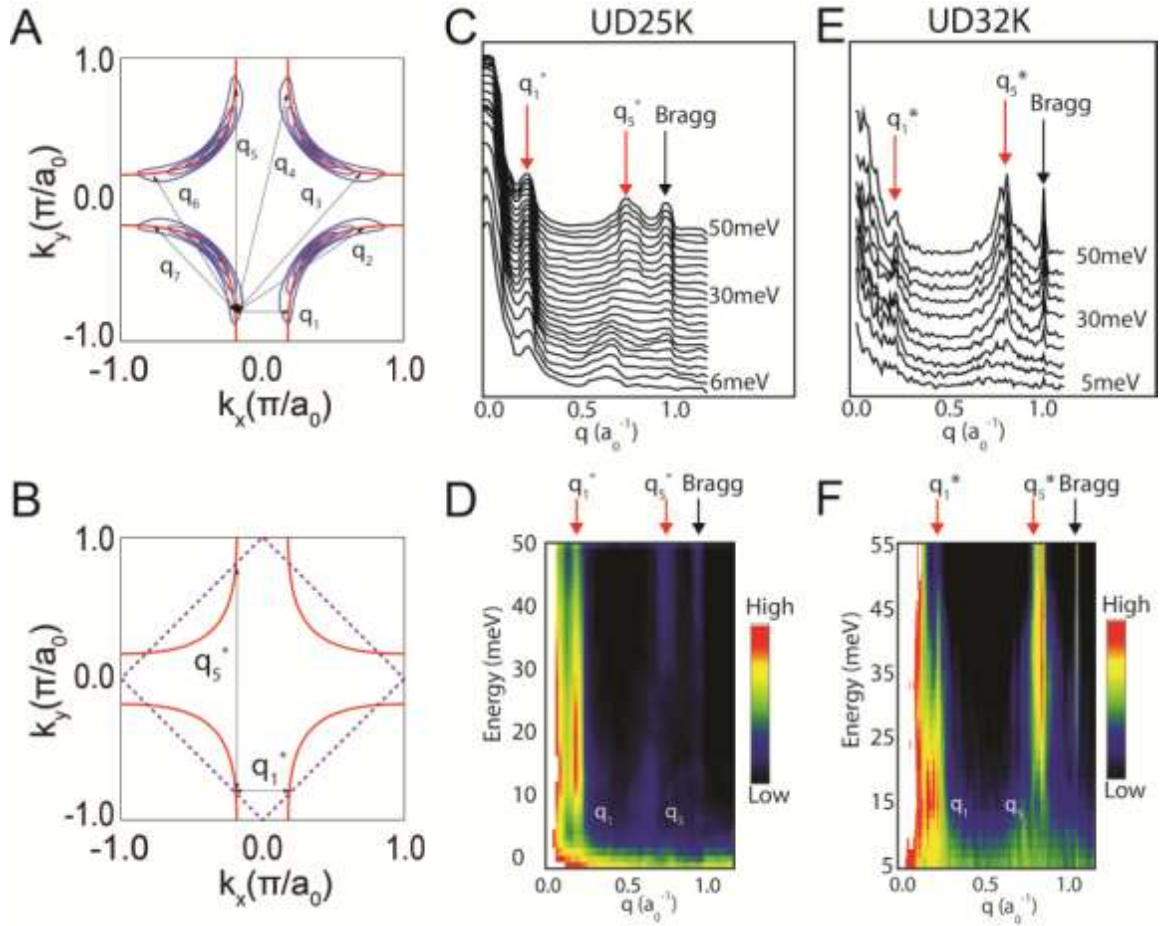

Figure S4 | Octet QPI extinction

A, Cartoon showing the seven dominant scattering wavevectors between the octet of high-DOS regions which disperse at energies within the superconducting gap. Schematized Fermi surface is shown in red, and schematized contours of constant energy are shown as thin blue lines. Both $q_1$ and $q_5$ appear along both the $q_x$ and $q_y$ axes. B, Cartoon showing the AFBZ (purple dashed line) where $q_1$ and $q_5$ stop dispersing and merge into the static high-energy wavevectors $q_1^*$ and $q_5^*$. C,D, Two views of the same linecut of $Z(q, E)$ from $(0, 0)$ to $(2\pi, 0)$ as a function of energy in UD25K. E,F, Two views of the linecut of $Z(q, E)$ from $(0, 0)$ to $(2\pi, 0)$ as a function of energy in UD32K. Dispersion of the octet QPI vectors $q_1$ and $q_5$ is observed. The octet vectors $q_1$ and $q_5$ disperse according to the octet model at low energies, then saturate at the static wavevectors $q_1^*$ and $q_5^*$ at energies above the superconducting gap.

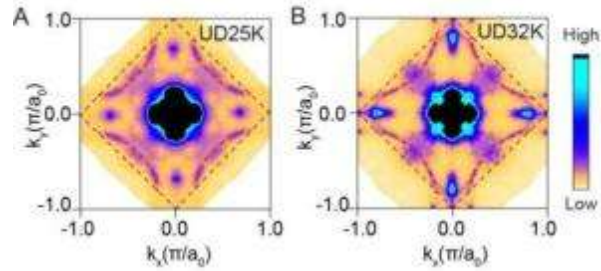

### Figure S5 | Energy-integrated QPI
A, B Sum of $Z(q,E)$ from 5mV to 25mV for UD25K and UD32K respectively. The trace of $q_4$ is highlighted with red dashed line. The trace of $q_4$ ($=2k_F$) mirrors the shape of the Fermi surface. Its cut-off at the AFBZ boundary (black dashed line) is clearly observed, as is the presence of antinodal QPI in the UD32K sample and its absence in the UD25K sample.

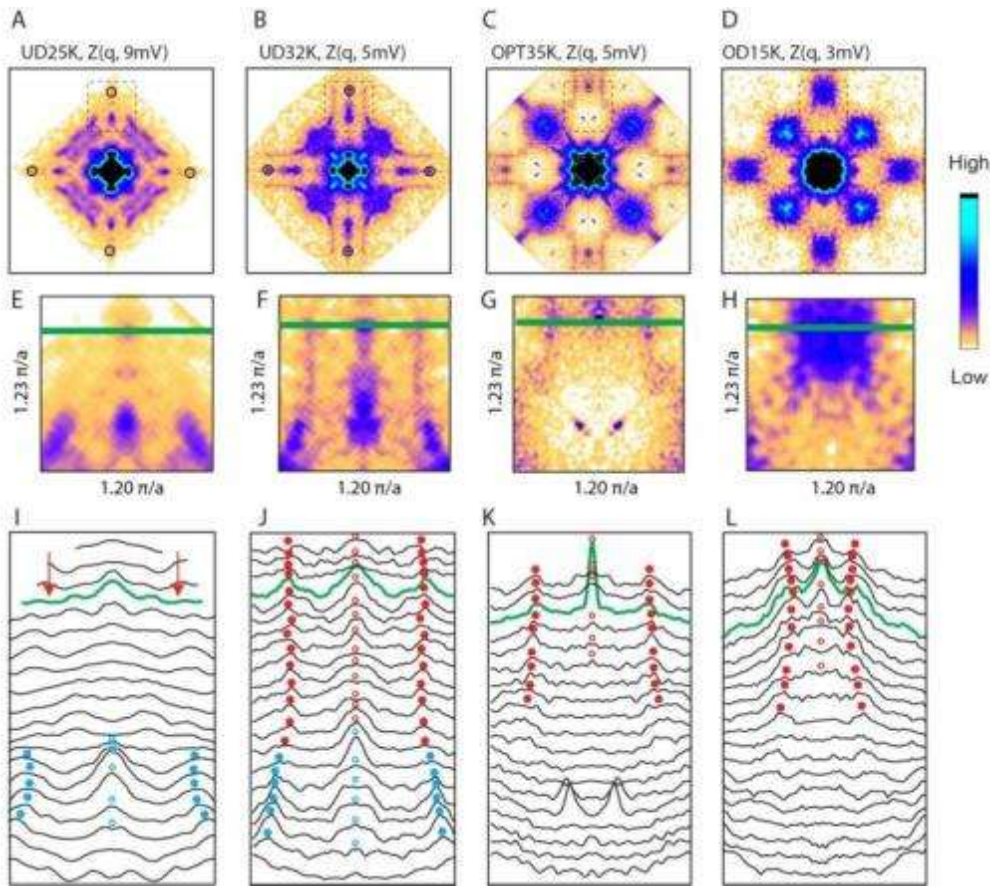

### Figure S6 | Antinodal QPI detail
A-D, $Z(q,E)$ as a function of doping A, UD25K at 9 mV; B, UD32K at 5 mV; C, OPT35K at 5 mV; D, OD15K at 3 mV. E-H, $Z(q,E)$ around the Bragg peak as a function of doping. I-L, waterfall plots of the data from E-H from the Bragg point (solid green line, top) towards the Γ point, through the antinodal region where the novel "triplet" QPI would appear. Octet QPI peaks ($q_4$ and $q_5$) are marked with blue dots (solid and open respectively). Antinodal QPI is marked with red dots. Arrows in the UD25K panel I show the locations where antinodal QPI would appear but is notably absent, consistent with a loss of the antinodal Fermi surface due to reconstruction.

## III. Antinodal QPI simulation

## A. Tight-binding Fermi surface

In optimally doped $Pb_{0.55}Bi_{1.5}Sr_{1.6}La_{0.4}CuO_{6+x}$ (with superconducting $T_c = 35$ K and pseudogap $T^* = 132 \pm 8$ K), a global fit to the ARPES energy dispersion curves measured at $T = 172$ K gives the following tight-binding model (15):

$$\varepsilon(k) = -2t_0 \left(\cos k_x + \cos k_y\right) - 4 t_1 \cos k_x \cos k_y - 2t_2 \left(\cos 2k_x + \cos 2k_y\right)$$
$$-4t_3 \left(\cos 2k_x \cos k_y + \cos k_x \cos 2k_y\right) - \varepsilon_0$$

$t_0 = 0.22; t_1 = -0.034315; t_2 = 0.035977; t_3 = -0.0071637; \varepsilon_0 = -0.24327$

In a separate work, doping-dependent tight-binding fits to the $Bi_2La_xSr_{2-x}CuO_{6+\delta}$ Fermi surface justify a rigid band model (48). We therefore fix $t_0$, $t_1$, $t_2$, and $t_3$, and adjust $\varepsilon_0$ to best match our QPI data for each of our four samples, as shown in Table S2 and Figure S7.

Table S2 | Tight-binding parameters

|  | $t_0$ | $t_1$ | $t_2$ | $t_3$ | $\varepsilon_0$ |
|---|---|---|---|---|---|
| UD25K | 0.22 | -0.034315 | 0.035977 | -0.0071637 | -0.15 |
| UD32K | 0.22 | -0.034315 | 0.035977 | -0.0071637 | -0.20 |
| OPT35K | 0.22 | -0.034315 | 0.035977 | -0.0071637 | -0.21 |
| OD15K | 0.22 | -0.034315 | 0.035977 | -0.0071637 | -0.25 |

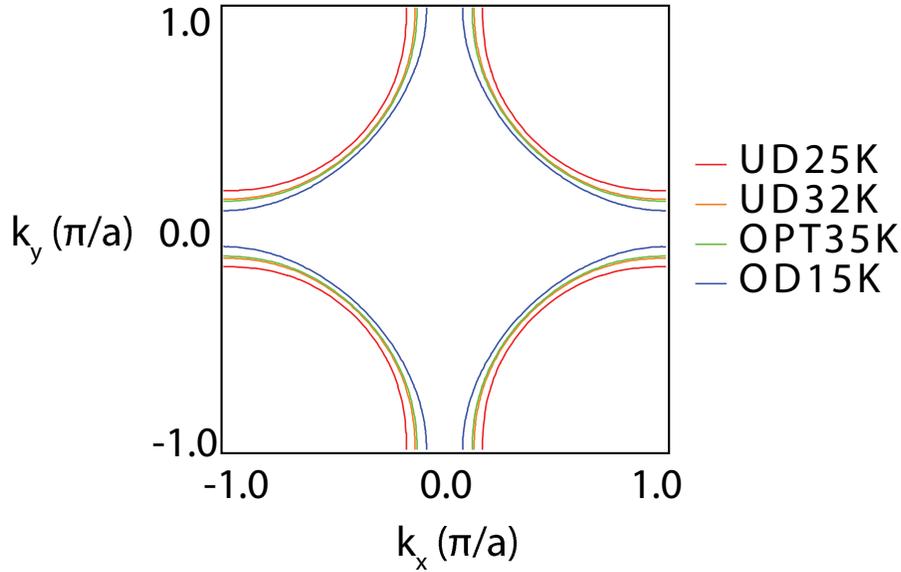

Figure S7 | Tight binding Fermi surfaces
Tight binding Fermi surface for four different doping levels.

## B. Model-free JDOS simulation

To demonstrate qualitatively that the antinodal QPI originates from the antinodal part of the FS, we

simply compute the autocorrelation(49) of the FS with ±30 meV broadening:

$$JDOS(q) = \int_{BZ} A(k)A(k+q)dk \quad \text{where} \quad A(k) = \begin{cases} 1, & -30 \text{ meV} \leq \varepsilon(k) \leq +30 \text{ meV} \\ 0, & \text{otherwise} \end{cases}$$

When the integral is restricted to regions of the BZ within 25° of the antinode for the OPT35K and OD15K samples, the JDOS simulation matches the QPI data well, as shown in main text Figs. 2K-L. However, we emphasize that the qualitative conclusion that the "triplet" QPI feature originates from the antinodal FS is robust to the value of the broadening energy and the angular cutoff used, as shown in Figure S8.

Because the antinodal autocorrelation emphasizes $q_4 = (2k_x, 2k_y)$, its shape should match the FS, scaled by a factor of 2. We display the JDOS and QPI data together to confirm the viability of our model, including the tight-binding fit, in Figure S9.

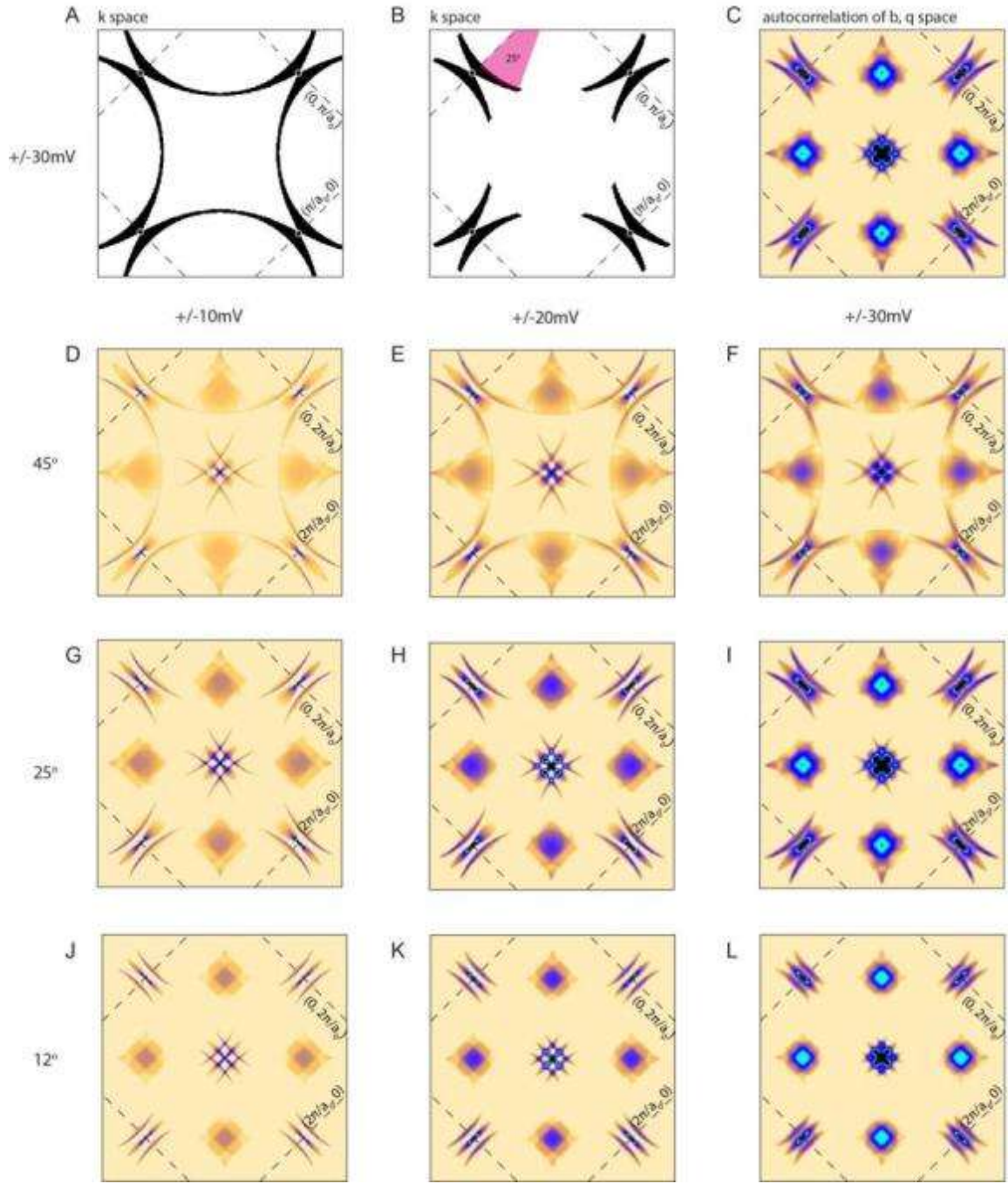

**Figure S8 | Antinodal JDOS simulation**
A, OD15K FS from tight-binding model, broadened by ±30 meV around the Fermi energy. B, The broadened FS is restricted to within 25° of the antinode. C, Autocorrelation of b (identical to main text Fig. 2l). Effects of varying the energy broadening and angular restriction, as indicated in column and row headings, are shown in D-L, demonstrating the robustness of the qualitative conclusion that the "triplet" QPI feature originates from the antinodal FS. Note that A-B are in $k$-space, whereas C-L are in $q$-space, so that the scale is doubled between these two sets of images.

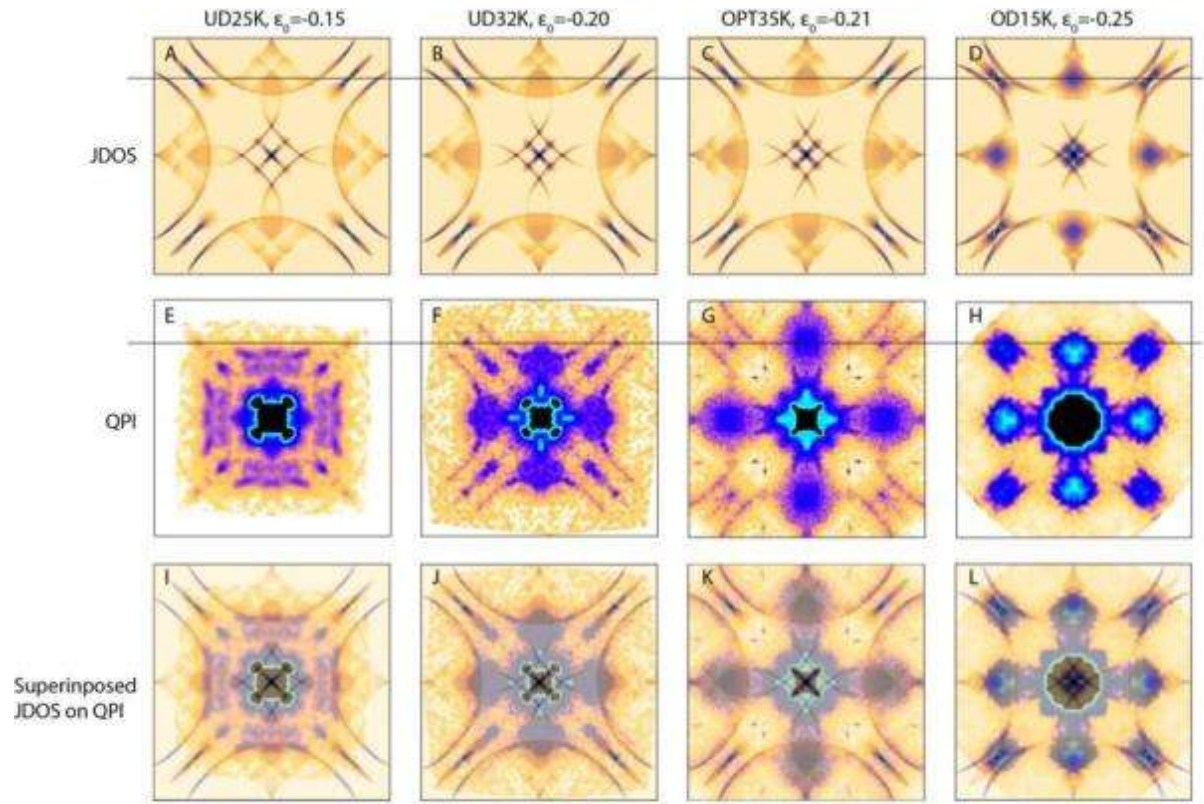

### Figure S9 | Tight-binding confirmation
A-D, JDOS simulations for all four samples, using the tight-binding parameters of Table S2, with broadening ±30 mV, and no angular cutoff. E-H, QPI data, $Z(\boldsymbol{q}, E)$, for all four samples at energies 9 mV, 5 mV, 5 mV, and 3 mV, respectively. I-L, superposition of calculated JDOS with experimental data. The excellent coincidence of $q_4 = (2k_x, 2k_y)$ in simulation and data justifies the tight-binding fit parameters.

## IV. Measures of hole count $p$

### A. Doping determination from $T_c$

Here we clarify several measures of doping. We define $x$ as the number of holes per Cu atom, and $n$ as the number of electrons per Cu atom. The "undoped" parent compound contains $x = 1$ hole (or equivalently $n = 1$ electron) per Cu site, corresponding to a half-filled band. However, strong Coulomb repulsion $U$ prevents double-electron occupancy of a single Cu site, so the holes are effectively "stuck" in a highly correlated Mott insulating state. To induce superconductivity, the material is chemically doped by adding $p \equiv x - 1$ holes per Cu site. These definitions are summarized in Table S3.

**Table S3 | Electron and hole concentration definitions**

| Symbol | Definition | Formula | Varies as… |
|---|---|---|---|
| $x$ | total number of holes per Cu atom | | Increases from $x = 1$ upon hole-doping the parent compound |
| $n$ | total number of electrons per Cu atom | $n \equiv 2 - x$ | Decreases from $n = 1$ upon hole-doping the parent compound |
| $p$ | number of additional doped holes per Cu atom | $p \equiv x - 1$ $= 1 - n$ | Increases from $p = 0$ upon hole-doping the parent compound |
| $n_{\text{Hall}}$ | mobile carrier concentration, as measured by Hall resistivity | | Increases from $n_{\text{Hall}} \sim 0$ in the parent compound to order $1 + p$ near optimal hole-doping |

Specific methods of hole doping vary across cuprates. For example, holes may be added by Sr substitution in La$_{2-z}$Sr$_z$CuO$_4$ (LSCO) or oxygen addition in Bi$_2$Sr$_2$CaCu$_2$O$_{8+d}$ (Bi2212); holes may be removed by La substitution in Bi$_2$Sr$_{2-y}$La$_y$CuO$_{6+d}$ (Bi2201) or oxygen removal in YBa$_2$Cu$_3$O$_{7-d}$ (YBCO). In properly annealed LSCO systems with stoichiometric oxygen content, $p$ may be identified directly with the Sr content, i.e. $p = z$. However, in cuprates where non-stoichiometric oxygen is involved, it is typically not possible to directly measure $p$.

A formula relating $T_c$ to $p$ in Bi2201 was developed by Ando *et al* based on a comparison of Hall measurements in several cuprates (*27*). The Hall number $n_{\text{Hall}}$ measures the mobile carrier concentration, and evidently must increase from 0 at $p = 0$ (where the $x = 1$ carriers are "stuck" rather than mobile at half-filling) to a number of order $x = p + 1$ as the material becomes a metal beyond $p \sim 0.16$. The details of this rapid increase in $n_{\text{Hall}}$ are not well understood, but empirically the addition of one doped hole in Bi2201 results in 7 additional mobile carriers over the range $0.1 < p < 0.15$ (*28*). Ando *et al* argued that the relationship between the room temperature value of $n_{\text{Hall}}$ (a directly measurable quantity) and $p$ (which is generally not directly measurable), is sufficiently universal across the cuprates to allow its use in defining $p$ in Bi2201 by comparison of its measured $n_{\text{Hall}}$ to that of LSCO where $p$ is known directly from the cation chemistry (*27*). This comparison gives an empirical relation for Bi2201

$$T_c = T_{c,\text{max}}[1 - 278(p - 0.16)^2]$$

which we have used to determine $p$ for our four samples from their measured $T_c$, in main text Fig. 2J.

For completeness, in Figure S11 we compare Ando's formula to a "universal" cuprate formula

$$T_c = T_{c,\text{max}}[1 - 82.6(p - 0.16)^2]$$

obtained from gravimetric analysis of Bi2212 by Presland *et al* (*50*). The Presland formula is based on two assumptions: first, that $p = 0.16$ corresponds to optimal doping; and second, each that oxygen contributes two holes. Because both Ando and Presland formulas are parabolas centered at $p = 0.16$, Ando's formula just linearly compresses the $p$-axis with respect to Presland's formula, so the Luttinger count vs. $p$ trend (Fig. 2J) remains linear for both definitions of $p$.

### B. Doping determination from Pseudogap

The hole doping can also be estimated from the pseudogap. Many formulas exist in the literature, several of which are compared in Figure S10.

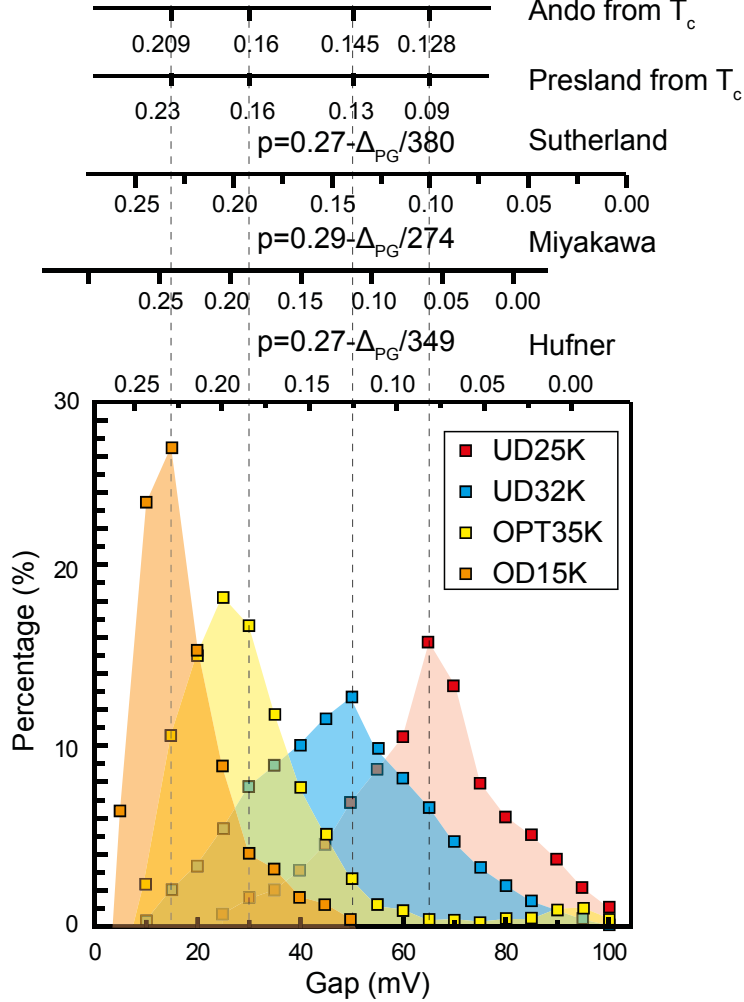

Figure S10 | Pseudogap distributions
Histograms of the $\Delta_{PG}$ distributions observed in the four samples studied. Dashed gray lines show $\Delta_{PG}$ for the average spectrum of each sample, which is slightly different from $\langle\Delta_{PG}\rangle$ computed from each individual spectrum then averaged. The measured $\Delta_{PG}$ (bottom axis) has been converted to doping using $p = 0.27 - \Delta_{PG}/349$ (51). Older variations on this formula (52, 53) along with conversions directly from bulk $T_c$ to doping (27, 50) give similar results, and in particular all five formulas agree that the UD32K sample (blue), in which both octet QPI extinction and antinodal QPI are observed, is centered around $p = 0.13 \pm 0.015$. Note that each histogram may have an artificial low-energy cutoff, as we are unable to distinguish $\Delta_{PG}$ when it becomes less than $\Delta_{SC}$. Because $\Delta_{PG} \gg \Delta_{SC}$ for all but the OD15K sample, this low energy cutoff is expected to affect the OD15K (orange) histogram only.

### C. Luttinger count

Here we clarify the Luttinger count on the vertical axis of main text Fig. 2J. Luttinger's theorem (54) says that $x = 2A_k/A_{BZ}$, where $A_k$ is the area of the hole pocket, which should be exactly $A_{BZ}/2$ in the undoped, half-filled parent compound. Luttinger's theorem is strictly a counting argument, which should apply to the total number of carriers in the band, whether or not they are mobile. Luttinger's theorem is commonly written in the cuprates in terms of $p$, as $1 + p = 2A_k/A_{BZ}$. However, symmetry breaking and possible zone folding complicate the situation (36, 55), so that in order to understand the small FS picture, it is simpler to count electrons, as explained in Table S4.

Table S4 | Large and small Fermi surface comparison

| Large Fermi surface | Small Fermi surface |
|---|---|
| (diagram: holes blue, electrons green) | (diagram: holes pink, electrons green, AFBZ dashed) |
| $A_{\text{green}} + A_{\text{blue}} = A_{\text{BZ}}$ | $A_{\text{green}} + A_{\text{pink}} = A_{\text{AFBZ}} = \frac{1}{2} A_{\text{BZ}}$ |
| $n_{\text{Lutt}} = \dfrac{2A_{\text{green}}}{A_{\text{BZ}}} = 2 - \dfrac{2A_{\text{blue}}}{A_{\text{BZ}}}$ | $n_{\text{Lutt}} = \dfrac{2A_{\text{green}}}{A_{\text{BZ}}} = 1 - \dfrac{2A_{\text{pink}}}{A_{\text{BZ}}}$ |
| $p_{\text{Lutt,large}} = 1 - n_{\text{Lutt}} = \dfrac{2A_{\text{blue}}}{A_{\text{BZ}}} - 1$ | $p_{\text{Lutt,small}} = 1 - n_{\text{Lutt}} = \dfrac{2A_{\text{pink}}}{A_{\text{BZ}}}$ |

When we combine $p_{\text{Lutt,small}}$ for the UD25K sample with $p_{\text{Lutt,large}}$ for the UD32K, OPT35K, and OD15K samples, we arrive at the expected linear relationship between $p_{\text{Lutt}}$ and $p_{\text{Ando}}$ (or $p_{\text{Presland}}$), as shown in Figure S11. This provides strong confirmation of the Fermi surface reconstruction occurring in underdoped Bi2201. The fact that the intercept and slope of our measured $p_{\text{Lutt}}$ vs. $p_{\text{Ando}}$ (or $p_{\text{Presland}}$) do not take the expected values of 0 and 1, respectively, might suggest the need to revisit the two assumptions of $p_{\text{opt}} = 0.16$ (which governs the intercept) and the number of holes per dopant (which governs the slope) in the $p_{\text{Ando}}$ and $p_{\text{Presland}}$ formulas. Another possibility which we considered and ruled out in Figure S8 is that corrections are needed to the tight-binding fit we use to parameterize our FS (15, 48).

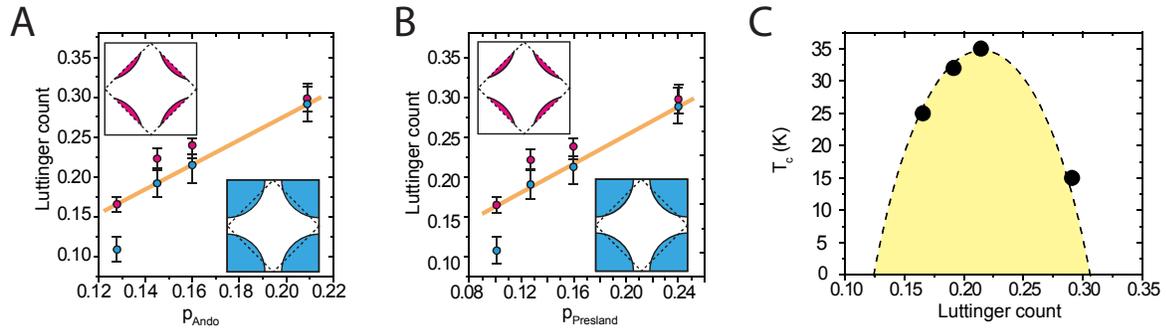

Figure S11 | Ando vs. Presland doping formula comparison
A, $p_{\text{Lutt}}$ vs. $p_{\text{Ando}}$ in the small (pink) and large (blue) FS scenarios. B, $p_{\text{Lutt}}$ vs. $p_{\text{Presland}}$ in the small (pink) and large (blue) FS scenarios. C, A new conversion from $T_c$ to hole doping $p$ using the Luttinger count determined here. We find a new optimal doping around $p_{\text{Lutt}} \sim 0.22$ which agrees well with the "universal" value of optimal doping $p_{\text{opt}} \sim 0.22$ (56) as determined by the thermopower at 290 K for several cuprates (57).

## V. Magnetic field dependence

### D. Momentum space

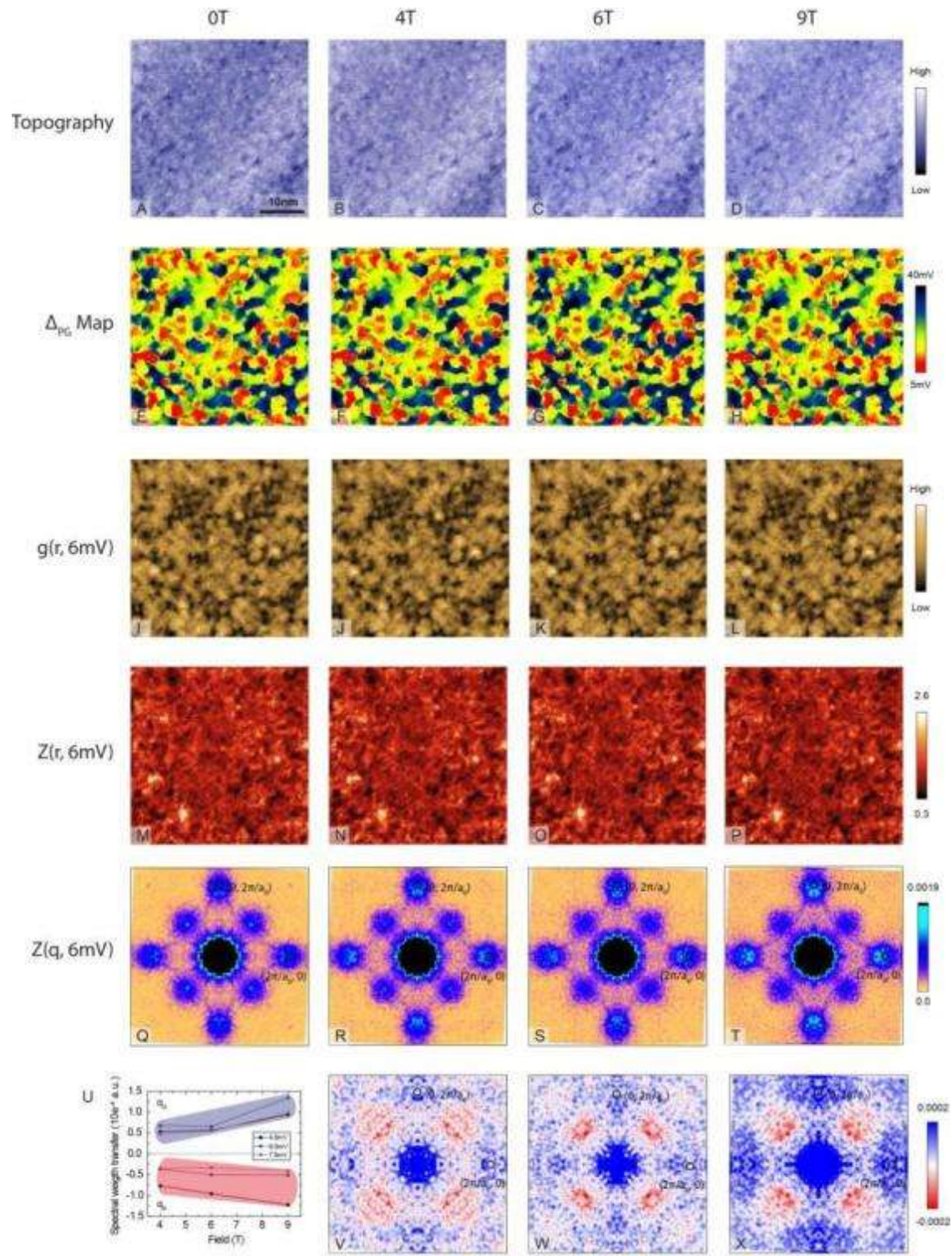

**Figure S12 | Magnetic field dependence**
A-D, $40 \times 40$ nm² topography of the same area of an OD15K sample, acquired at 0T, 4T, 6T, and 9T. The topographies appear identical, proving that the STM tip did not change between these maps, which were acquired over the course of a week. E-H, Pseudogap maps for the same field of view, calculated independently from data in all 4 fields, demonstrating the field-independence of $\Delta_{PG}(r)$. I-L, Raw $g(r, 6\,\text{mV})$ from all 4 fields, again demonstrating the lack of tip change. M-P, Ratio maps $Z(r, 6\,\text{mV})$ for all 4 fields. Q-T, Symmetrized and lightly smoothed $Z(q, 6\,\text{mV})$ for all 4 fields, showing the antinodal QPI. U-X, Field-induced scattering. V, $Z(q, 6\,\text{mV}, 4\,\text{T}) - Z(q, 6\,\text{mV}, 0\,\text{T})$; W, $Z(q, 6\,\text{mV}, 6\,\text{T}) - Z(q, 6\,\text{mV}, 0\,\text{T})$; X, $Z(q, 6\,\text{mV}, 9\,\text{T}) - Z(q, 6\,\text{mV}, 0\,\text{T})$; U, Trends showing the $B$-dependence of low-energy scattering. Sign-preserving scattering ($q_A$) is enhanced while sign-changing scattering ($q_B$) appears relatively suppressed as $B$ is increased.

We employ magnetic-field-dependent QPI imaging as a phase-sensitive probe of the $d$-wave order parameter. We investigate OD15K Bi2201, an ideal doping level in which to explore the interplay between superconductivity ($\Delta_{SC} = 6\,\text{mV}$) (*14*) with the pseudogap ($\langle\Delta_{PG}\rangle \sim 15\,\text{mV}$, Figure S10) far from the "intertwining" (*58*) orders which have recently proliferated in underdoped materials. We acquire differential conductance maps over identical $40 \times 40$ nm² areas at four different applied $B$-fields. The ability to track the same field of view (FOV) is confirmed by the identical topographic images in Figure S12A-D. The pseudogap energy $\Delta_{PG}$ is independent of $B$-field up to 9T, as demonstrated by the independently-computed but indistinguishable $\Delta_{PG}$-maps at the four fields in Figure S12E-H. The density of states (Figure S12I-L) and ratio maps (Figure S12M-P) are also indistinguishable by eye at the four fields, but Fourier transformations (Figure S12Q-T) reveal subtle differences (Figure S12V-X), whose trends are quantified in Figure S12U, and explained as follows.

In a superconductor with a complex order parameter $\Delta(k)$, elastic scattering processes between states $k_i$ and $k_f$ can be grouped according to whether $\Delta(k_i)$ and $\Delta(k_f)$ have the same or opposite sign. Wang & Lee first drew attention to the role of the superconducting coherence factors in suppressing sign-reversing scattering from magnetic impurities while suppressing sign-preserving scattering from scalar impurities (*59*). Pereg-Barnea & Franz pointed out that such coherence factors could be used to distinguish Bogoliubov quasiparticle scattering from other scattering processes unrelated to superconductivity (*60*). The challenge in applying these ideas arises from the difficulty in determining which types of impurities (magnetic vs. non-magnetic) are actually present in a given sample.

A breakthrough idea predicted that magnetic field could be used to increase magnetic scattering, thus enhancing sign-preserving scattering processes and suppressing sign-reversing scattering processes (*31*). This prediction was born out in Na$_x$Ca$_{2-x}$CuO$_2$Cl$_2$ (Na-CCOC) where all seven $q_i$'s behave exactly as expected: sign-preserving scattering vectors ($q_1, q_4, q_5$) are enhanced whereas sign-reversing scattering vectors ($q_2, q_3, q_6, q_7$) are suppressed by magnetic field. Furthermore, the field-enhanced scattering of $q_1, q_4$, and $q_5$ was most prominent in the regions immediately around the magnetic vortex cores (*30*). Additional theoretical calculations by Maltseva & Coleman pointed out that the ~10nm vortex cores in Na-CCOC were likely too large to enhance the short wavelength (large $q$) scattering processes. They concluded that the dominant field-enhanced scattering arises not from the vortices themselves, but rather from impurities within the pinned vortex cores (*61*).

These ideas were subsequently applied to the new Fe-based superconductors, with wholly different pairing symmetry. Magnetic field was again shown to enhance some scattering processes and suppress others, from which a sign-changing $s_\pm$ order parameter was inferred in FeSe$_{0.4}$Te$_{0.6}$ (*62*). However, in this case the field-enhanced scattering was not localized around the vortices, but was homogeneous throughout the sample. Follow-up theoretical work clarified that in Pauli-limited superconductors the Zeeman effect can enhance the sign-preserving scattering from impurities, without the need for pinned vortices (*63*).

In our present experiment on OD15K Bi2201, we image no static vortices, consistent with the expected vortex liquid phase (*64, 65*). From previous work on OD31K Bi2201, where static vortices of diameter ~10nm were imaged at 4T (*66*), we infer that the size of the moving vortex cores may be even larger in our OD15K sample due to the smaller superconducting gap. Therefore, similar to Na-CCOC, the field-dependence of the short wavelength $q_A$ and $q_B$ in our experiment may arise from scattering off impurities which spend much of their time within the moving large vortex cores.

## E. Real space

As discussed in the main text, we separate the superconducting component of the density of states by computing $S(\mathbf{r}, E) = g(\mathbf{r}, E, 0\text{ T}) - g(\mathbf{r}, E, 9\text{ T})$ to cancel the $B$-independent (pseudogap related) component. We emphasize that $\Delta_{\text{PG}}(\mathbf{r})$ is field-independent, as demonstrated in Figure S12E-H. The resulting $S(\mathbf{r}, E)$ are binned by the local $\Delta_{\text{PG}}(\mathbf{r})$, then averaged over all locations within each bin to produce a single average spectrum $\langle S(E) \rangle|_{\Delta_{\text{PG}}}$ for each value of $\Delta_{\text{PG}}$. This set of spectra is reproduced here in Figure S13A. The subtraction emphasizes the superconducting gap $\Delta_{\text{SC}}$ which was previously obscured by the dominant $B$-independent pseudogap $\Delta_{\text{PG}}$. We note that despite a factor of five change in the value of $\Delta_{\text{PG}}$, there is little change in the value of $\Delta_{\text{SC}} \sim 6$ meV throughout these spectra.

Dynes noted that the superconducting density of states, $\rho_s(E) = \frac{|E|}{\sqrt{E^2 - \Delta^2}}$, could be generalized to take into account a finite quasiparticle lifetime by writing

$$\rho_s(E, \Gamma) = Re\left(\frac{E - i\Gamma}{\sqrt{(E - i\Gamma)^2 - \Delta^2}}\right)$$

where $\Gamma$ is the inverse quasiparticle lifetime (*67*). In Figure S13B, we simulate the effect of increasing $\Gamma$ on a $d$-wave superconducting spectrum with a fixed gap $\Delta_{\text{SC}} = 6$ meV. We note that the broadening increases the *apparent* gap width (defined by the coherence peak maxima), despite the fixed $\Delta_{\text{SC}} = 6$ meV. This simulation matches well with the data in Figure S13A.

It is important to check for setup condition artifacts in subtracted spectra. In the main text, all the spectra were normalized to a constant $g(40\text{ mV})$. In Figure S14 we check several different normalizations and find qualitative consistency for all normalizations.

Both division (*14*) and subtraction (*33*) of superconducting and normal-state spectra have been used in cuprate studies to capture the superconductivity signal. For completeness, we show in Figure S15 the comparison between $g(E, 0\text{T})/g(E, 9\text{T})$ and $g(E, 0\text{T}) - g(E, 9\text{T})$. The suppression of the coherence peak by the pseudogap is qualitatively similar in both methods.

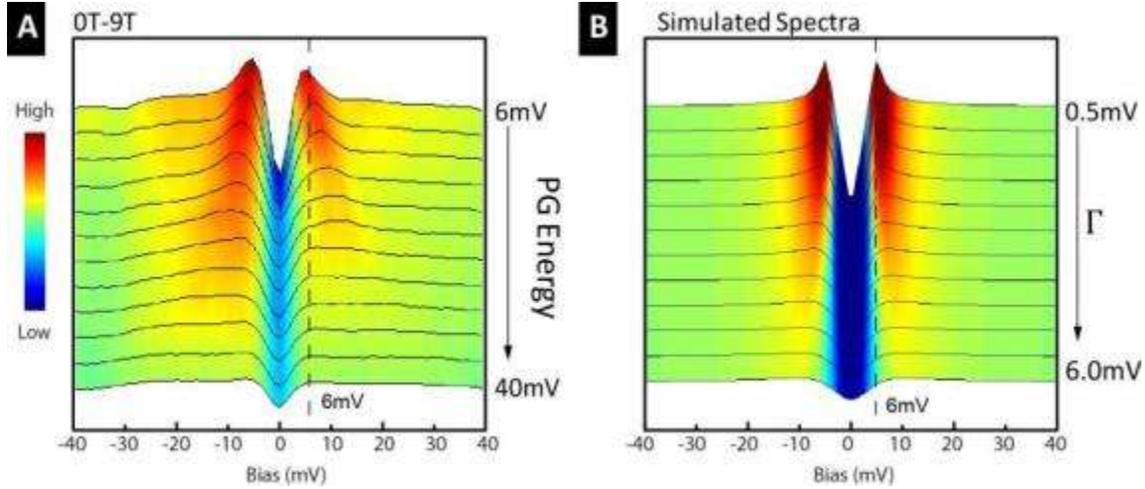

Figure S13 | Simulated spectral broadening
A, A set of average spectra $\langle S(E) \rangle|_{\Delta_{PG}}$ for values of $\Delta_{PG}$ ranging from 6 mV to 40 mV, offset for clarity. The pseudogap behaves like a broadening factor: increased local $\Delta_{PG}$ has the effect of increasing $\Gamma$. B, Simulated superconducting density of states with 6 mV $d$-wave gap, broadened by the Dynes formula within increasing $\Gamma$. Note that in both the data and the simulated spectra, $\Delta_{SC}$ appears to widen with increasing $\Delta_{PG}$ and $\Gamma$ respectively, although the actual energy of $\Delta_{SC}$ is constant. This can be simply understood as the effect of smoothing a sharp peak with different background on both sides: as the peak is smoothed, it will appear to move towards the side with higher background.

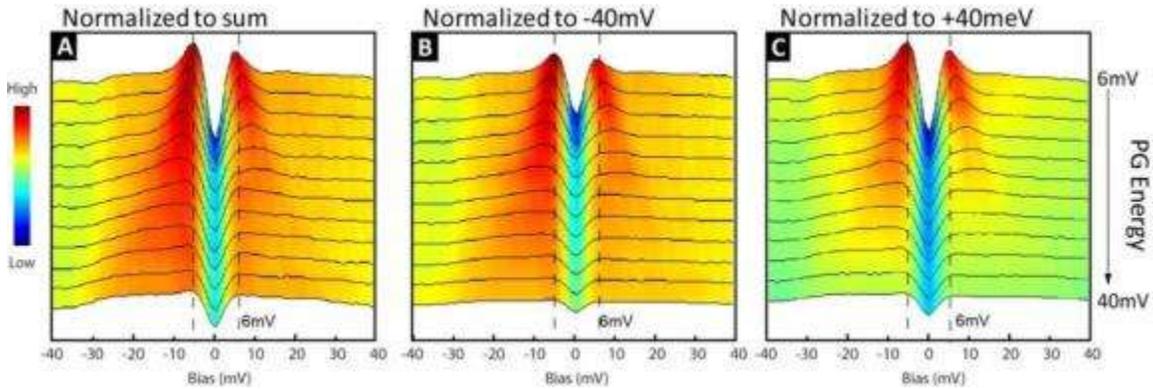

Figure S14 | Setup condition comparison
The magnetic field induced weight transfer $g(E, 0T) - g(E, 9T)$ versus pseudogap energy is shown with spectra normalized to A, the full integral of the spectrum $\int_{-40}^{+40} \frac{dI}{dV}(V)dV$; B, the value of $\frac{dI}{dV}(+40 \text{ mV})$; and C, the value of $\frac{dI}{dV}(-40 \text{ mV})$. The latter is reproduced from the main text Fig. 4B.

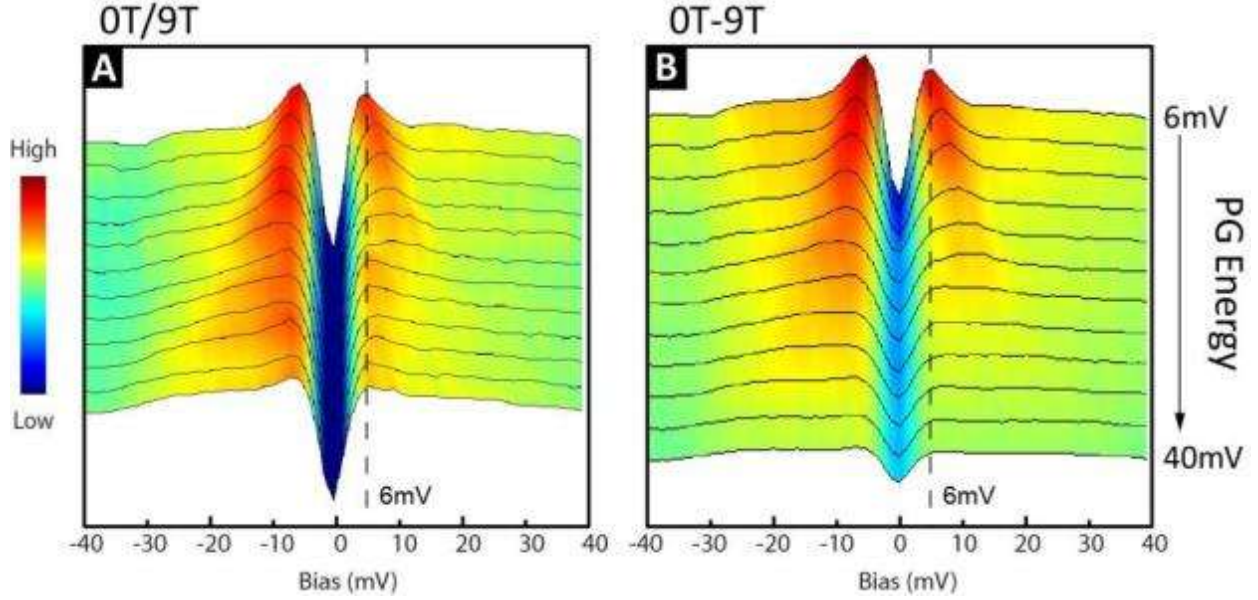

**Figure S15 | Normalization comparison: division vs. subtraction**
A, Binned and averaged spectra $g(E, 0T)/g(E, 9T)$ with varying $\Delta_{PG}$. B, Binned and averaged spectra of $g(E, 0T) - g(E, 9T)$ with varying $\Delta_{PG}$. The latter is reproduced from the main text Fig. 4B.

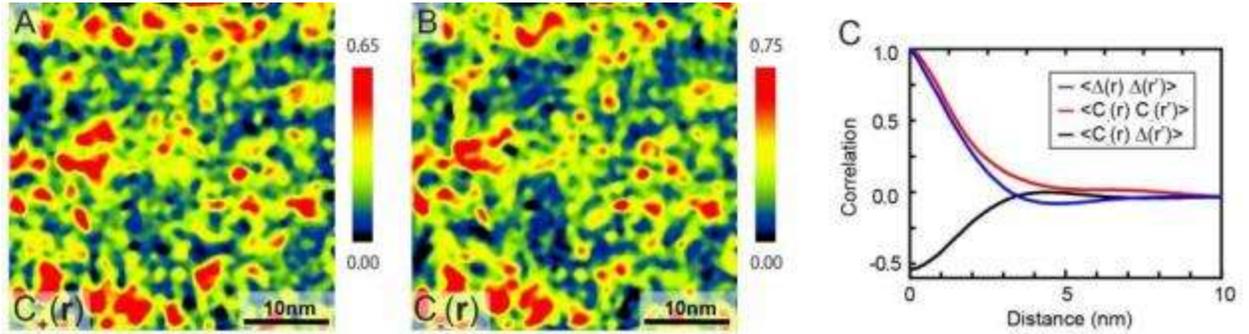

**Figure S16 | Coherence peak weight**
A, $C_+(r)$ map defined as the difference between the positive energy coherence peak amplitude and the zero bias conductance, $S(r, 6\,mV) - S(r, 0\,mV)$. Reproduced from main text Fig. 4d. B, $C_-(r)$ map defined as the difference between the negative energy coherence peak amplitude and the zero bias conductance, $S(r, -6\,mV) - S(r, 0\,mV)$. C, Cross-correlations involving $C_-(r)$ and the PG magnitude $\Delta_{PG}(r)$. Similar to the results for $C_+(r)$ in Fig. 4f, $C_-(r)$ anti-correlates with the pseudogap energy $\Delta_{PG}(r)$, demonstrating the pseudogap correlates with decoherence of superconductivity. The pseudogap may cause the superconducting decoherence, or both PG and SC decoherence may be caused by a third agent such as disorder (*68–70*).

## VI. Charge order in Bi-2201

We observe incommensurate charge order in all doping levels studied in this work. It is important to note that such charge order exists both below and above the doping level of the Fermi surface reconstruction we report here. This result, together with the ubiquity of the spectroscopic pseudogap presented in Figure 1B, is in good agreement with previously proposed phase diagrams (*22*)(*21*)(*17*)(*23*). One possible explanation of our results is thus that a fluctuating charge density modulation causes the pseudogap while a static charge density wave exists over a narrower doping range starting at the Fermi surface reconstruction. Alternatively, since charge order persists above the transition, this may be a 'first topological transition' restoring only part of the large Fermi surface, similar to that found in electron-doped cuprates (71, 72). Indeed, Storey et al. (73) found evidence for the appearance of an electron pocket in Bi2201 at essentially the same doping.

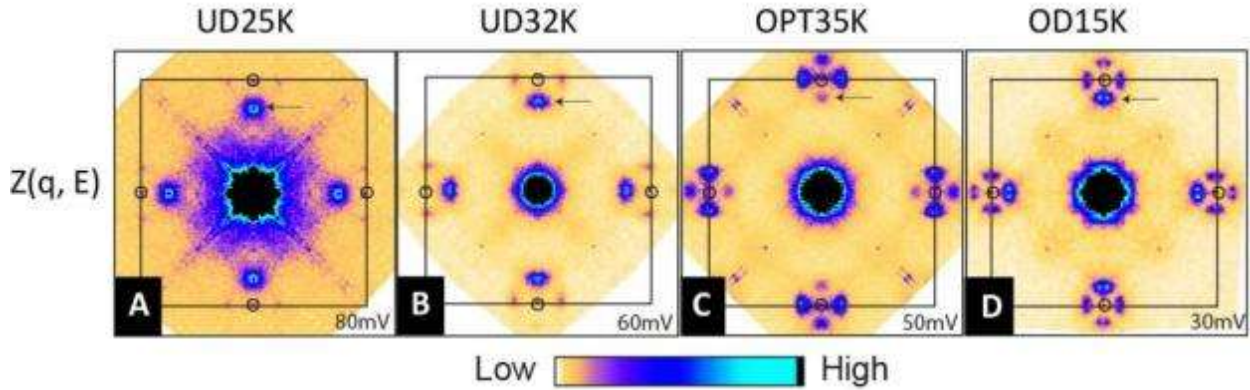

Figure S17 | Charge order in Z(q, E)
A-D, $Z(q, E)$ maps of the four samples at energies near the pseudogap energy. The charge order peak $\sim(\frac{3}{4} \cdot \frac{2\pi}{a_0}, 0)$ exists in all doping levels below and above the Fermi surface reconstruction reported in this work. Combining this result with the spectroscopic pseudogap (Figure 1B) shows good agreement with the phase diagram proposed by (22)(21)(17)(23) where the charge order is related to the pseudogap phase.

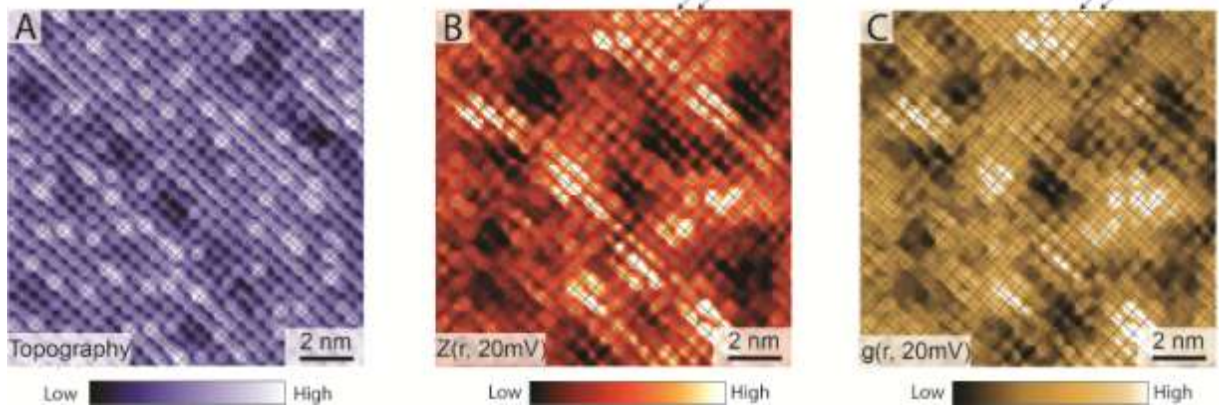

Figure S18 | Charge order in real space in OD15K
A, 10 nm by 10 nm topography with atomic resolution on our most overdoped sample OD15K. B-C, $Z(r, 20\text{mV})$ and $g(r, 20\text{mV})$ map in the same field of view overlaid with the atomic lattice grid from A. A unidirectional real space charge modulation is clearly observed, similar to the observation in Bi2212 (22)(21)(17)(23). Note that the modulations observed in B and C are distinct from atomic topography, which is highlighted by the overlaid grid. The q-space map of the same sample is presented in Figure S17D